\documentclass{WileyMSP-template}
\usepackage{soul}
\usepackage{xcolor}
\usepackage{mathtools}

\begin{document}

\pagestyle{fancy}
\rhead{\includegraphics[width=2.5cm]{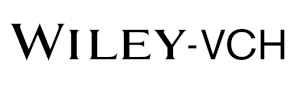}}

\title{Q-switched Mode-locking in Er-doped ZBLAN Fibre Lasers using Carbon Nanotube Saturable Absorber and GaSb-based SESAM}

\maketitle

\author{Boris Perminov*}
\author{Aram Mkrtchyan}
\author{Yuriy Gladush}
\author{Dmitry V. Krasnikov}
\author{Albert G. Nasibulin}
\author{Maria Chernysheva*}

% Dedication

%\dedication{Optional dedication here. If no dedication is required, please leave blank}

% Affiliations: Please provide adacemic titles (Prof. or Dr.) for all authors where applicable, and include an institutional email address for all corresponding authors
\begin{affiliations}
B. Perminov, Dr M. Chernysheva\\
Leibniz Institute of Photonics Technology, Albert-Einstein-Str. 9, 07745 Jena, Germany\\
Email Address: Boris.Perminov@leibniz-ipht.de, Maria.Chernysheva@leibniz-ipht.de

Dr A. Mkrtchyan, Prof Y. Gladush, Prof D. V. Krasnikov, Prof A. G. Nasibulin\\
Skolkovo Institute of Science and Technology, 30 Bolshoy Boulevard 1, Moscow, 121205, Russia

\end{affiliations}

\keywords{fibre laser, Q-switched mode-locking, Saturable absorbers, carbon nanotubes}

\begin{abstract}

Mid-infrared fibre lasers are crucial for applications in spectroscopy, medical diagnostics, and environmental sensing, owing to their ability to interact with fundamental molecular vibrational bands. However, achieving stable ultrafast pulse generation in this spectral range remains challenging due to the limited availability of robust saturable absorbers. For the first time, we demonstrate Q-switched mode-locking in an all-fibre Er-doped ZBLAN laser employing an aerosol-synthesised carbon nanotube film. Furthermore, we compare the laser performance with pulse generation using a state-of-the-art GaSb-based SESAM in an identical cavity design. The carbon nanotube saturable absorber enables pulse generation with a minimum duration of 1.32~$\mu$s and a pulse energy of 1.4~$\mu$J at an average output power of 63.1~mW. In contrast, the SESAM-based laser produces 560-ns pulses with a pulse energy reaching 4.42~$\mu$J and an average power of 138~mW. These results provide new insights into the interplay between saturable absorber properties and mid-IR fibre laser performance, paving the way for next-generation compact ultrafast sources for scientific and industrial applications.
\end{abstract}

\section{Introduction}
\justify
Mid-infrared (mid-IR) laser technologies play a vital role in advancing existing and paving the way for emerging opportunities in molecular spectroscopy, medical diagnostics, environmental monitoring, and defence systems. The mid-IR vibrational absorption bands of various gases, biomolecules, and pollutants enable non-invasive assessment of the chemical structure and biochemical content~\cite{thorpe2006broadband,schwartz2019remote, bareza2020mid}. However, the mid-IR spectral range from 2.5 to 4.5~$\mu$m has not yet reached industrial maturity due to the limited availability of light sources, such as quantum cascade lasers, crystal-based lasers, and optical parametric systems, each with distinct limitations~\cite{grebnev2024fluoride}.

Fluoride fibres hold great potential to apply effective, compact, cost-efficient, and alignment-free all-fibre laser systems to the mid-IR range. Er$^{3+}$-doped fluoride fibres, including fluorozirconate (such as ZBLAN with a ZrF$_4$-BaF$_2$-LaF$_3$-AlF$_3$-NaF glass matrix) and fluoroindate fibres, have emerged as promising gain media for efficient emission around 2.8~$\mu$m due to $^{4}$I$_{11/2}$ $\rightarrow$ $^{4}$I$_{13/2}$ transition in Er$^{3+}$ ions \cite{duval2015femtosecond,wang2022wavelength}. Continuous-wave Er-doped fluoride fibre lasers and amplifier systems have achieved 40~W output powers~\cite{aydin2018towards}, which could be further increased by switching to quasi-CW generation~\cite{newburgh2021power}.

Despite high application potential, the development of compact, high-performance ultrafast fibre lasers remains a challenge. A significant limitation is the lack of robust saturable absorbers or fast modulators tailored for this spectral region. Moreover, demonstrated all-fibre lasers in the mid-IR range are rare, with only a few reported examples~\cite{ma2019review}. Currently, the application of nonlinear polarisation evolution is the most established technique for pulse generation. However, it relies on free-space light propagation through bulk components, such as wave plates, polarisation beam splitters, and isolators~\cite{duval2015femtosecond,bawden2021ultrafast,yu2022average}. 
Pulse durations below 150~fs have been achieved directly from the oscillator~\cite{gu2020generation}. Furthermore, with the nonlinear application in Er$^{3+}$-doped ZBLAN fibre, the pulse could be further compressed below sub-100 fs duration due to the self-phase modulation broadening of the spectrum, reaching average power of 2.4~W~\cite{cui2021generation}.
Additionally, the overlap of the laser spectrum with the absorption bands of the atmospheric gases, mentioned as beneficial for sensing applications, can deteriorate beam quality and introduce spectral filtering, ultimately limiting the generated spectral bandwidth and pulse duration~\cite{majewski2021picosecond}.

Among material saturable absorbers, semiconductor saturable absorber mirrors (SESAMs) based on GaAs heterostructures have dominated ultrafast laser technology in the near-infrared region~\cite{keller1996semiconductor}. These devices feature monolithic designs with distributed Bragg reflectors (DBR), enabling low non-saturable losses and high damage thresholds. However, their performance declines at wavelengths beyond 2.5~$\mu$m due to bandgap limitations. Recent advances in GaSb-based SESAMs have extended their utility towards 3~$\mu$m~\cite{alaydin2022bandgap,normani20242}. The GaSb-SESAM leverages type-I quantum wells and a monolithic AlSb/GaSb DBR structure, providing a broad bandgap spanning from 2.65 to 2.97~$\mu$m (corresponding to 0.72~eV) and ultrafast recovery ($<$1~ps). The recent work reported a 30-ps pulse generation in GaSb-SESAM mode-locked Er-doped fibre laser with 190~mW average power~\cite{normani20242}. Still, the demonstrated laser system relied on free-space optical elements, such as a diffraction grating, a couple of dichroic mirrors to direct the pump light and a polariser~\cite{paradis2023ultrafast,qin2022semiconductor,luo2021high}. 

Alternatively, nanomaterials, such as graphene, transition metal dichalcogenides, and carbon nanotubes (CNTs), have been extensively studied for establishing ultrashort pulse operation in various fluoride fibre laser systems. For instance, MXene-based SAs have facilitated Q-switched generation with an average output power of more than 970~mW and pulse durations of approximately 1~$\mu$s pulses in Er-doped ZBLAN fibre laser~\cite{wei2020mxene}. Additionally, a linear configuration of all-fibre laser that operated both in Q-switched and mode-locked regime at 2.8~$\mu$m using black phosphorus has been reported. These results underscore the potential of 2D materials for mid-IR applications \cite{qin20182}. In contrast to 2D nanomaterials, CNTs exhibit wavelength-selective saturable absorption due to their bandgaps being influenced by their diameter and chirality of the nanotubes~\cite{hasan2009nanotube,chernysheva2017carbon,sobon2017cnt}. Single- and double-walled CNTs have already demonstrated high efficiency in ultrashort pulse generation across the 1300--2400~nm spectral range, achieved by tailoring the nanotube diameter to precisely adjust the bandgap of semiconducting nanotubes, ensuring resonance with the laser emission wavelength~\cite{kelleher2009nanosecond,khegai2018bismuth,set2004laser,chernysheva2012thulium,kozlova2025self}. However, the application of CNTs as saturable absorbers in the mid-IR range encounters several challenges, such as an increase in wall defects resulting from larger nanotube diameters. Overall, the limited modulation depth at longer operating wavelengths and thermal instability at high power levels have restricted the industrial use of nanomaterial-based saturable absorbers.

This work reports on the comparison of Q-switched mode-locking in linear Er-doped ZBLAN fibre lasers operating at 2.8~$\mu$m, employing two saturable absorbers: a GaSb-based SESAM and an aerosol-synthesised CNT film. Our results provide valuable insight into the interplay between saturable absorber properties and mid-IR fiber laser dynamics, guiding future advancements toward stable mode-locking. 

\section{Experimental results}

\subsection{Passive Q-switching using a CNT saturable absorber}

\begin{figure}[bp]
 \centering
  \includegraphics[width=0.98\linewidth]{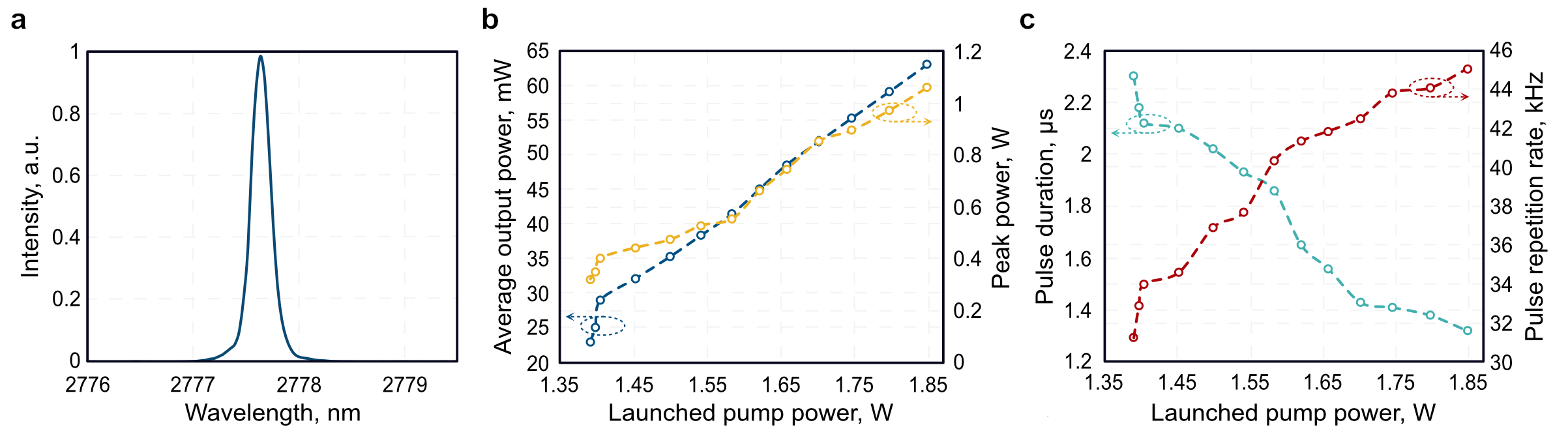}
  \caption{Output parameters of CNT Q-switched Er-doped ZBLAN fibre laser: (a) Normalised output spectrum; variations of (b) output average and peak powers and (c) pulse repetition rate and duration with output power.}
  \label{fig:CNToutput}
\end{figure}

We have successfully demonstrated Q-switched laser operation at 2777.5 nm using a carbon nanotube (CNT) saturable absorber, achieving a spectral bandwidth of 0.2 nm, as depicted in Fig.\ref{fig:CNToutput}(a). The laser’s performance was comprehensively analysed by varying the pump power, with key output parameters—average power, pulse energy, pulse duration, and repetition rate—summarised in Fig.~\ref{fig:CNToutput}. Stable Q-switching was observed at a threshold pump power of 1.37~W. At the initiation of Q-switching, the laser produces pulses at a repetition rate of 31.25~kHz, with a pulse duration of 2.3~$\mu$s and an average output power of 23~mW. As the pump power increased, the average output power rose monotonically, reaching 63~mW with 8.3\% slope efficiency. Concurrently, the pulse duration shortened from 1.32~$\mu$s, and the pulse repetition rate increased from 45~kHz, as illustrated in Fig.\ref{fig:CNToutput}(c), following inherent dynamics of passively Q-switched lasers, since increased pumping rates accelerate inversion and saturation dynamics in the gain medium and SA. A maximum pulse energy reached 1.4~$\mu$J, as shown in Fig.~\ref{fig:CNToutput}(b), corresponding to 1.06~W peak power.

\begin{figure}[!]
 \centering
  \includegraphics[width=0.48\linewidth]{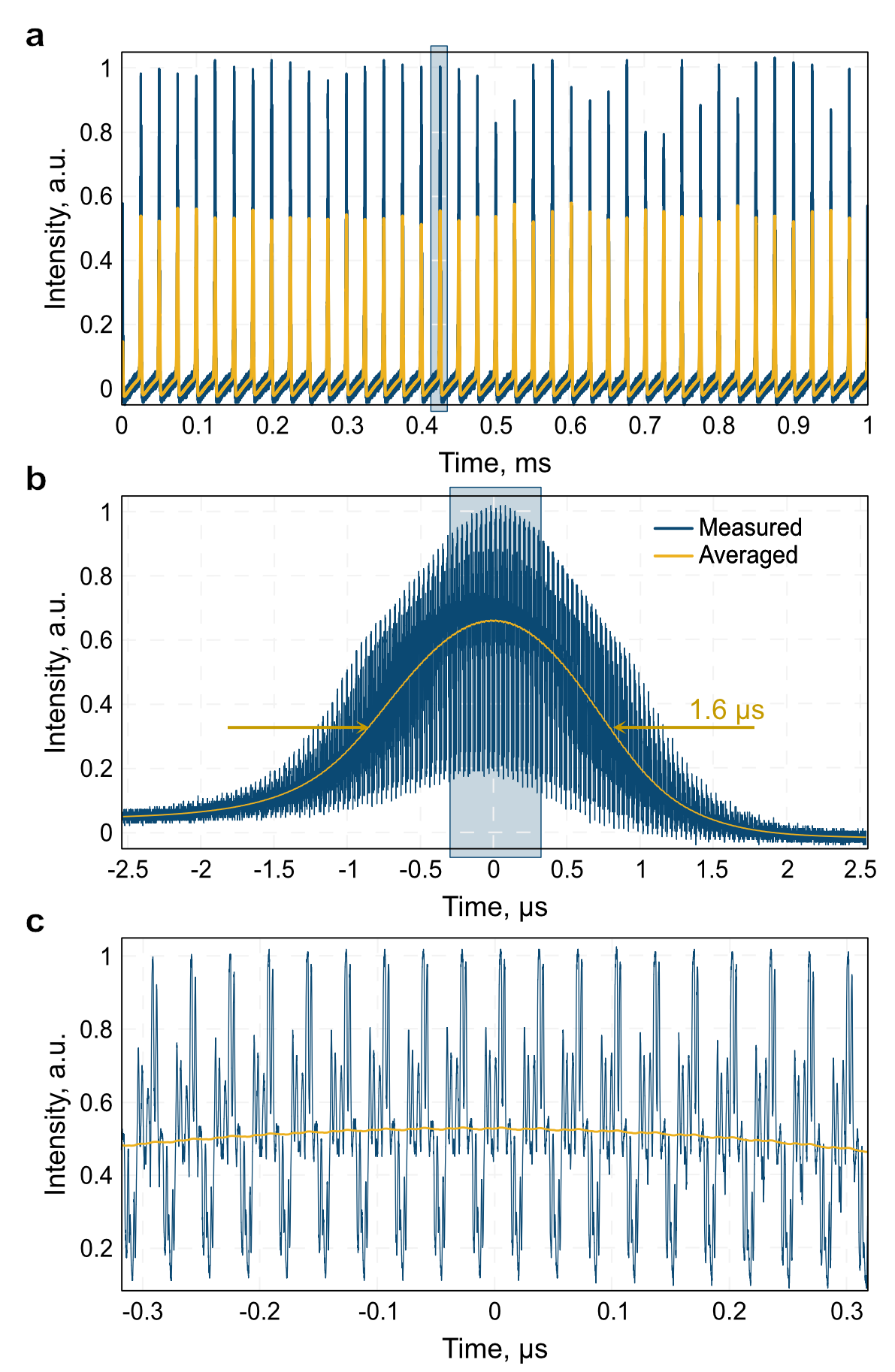}
  \caption{Characteristic output pulse train of CNT Q-switched Er-doped ZBLAN fibre laser : measured (blue plot) and averaged (yellow plot) (a) at 1-ms time span; (b) zoomed-in individual Q-switched pulse; (c) Zoomed-in oscillations at laser fundamental mode-locking repetition rate}
  \label{fig:CNTpulse}
\end{figure}

The oscilloscope trace of the pulse train presented in Fig.~\ref{fig:CNTpulse}(a) reveals pronounced intensity fluctuations with a standard deviation of nearly 6.45\% over the recorded 1~ms time span. Such intensity modulation is caused by rapidly fluctuating temporal patterns, which vary from pulse to pulse, as shown in Fig.~\ref{fig:CNTpulse}(b). As seen in the zoomed-in oscilloscope trace, these mode-locking-like oscillations exhibit a regular temporal spacing of 35.7~ns (Fig.~\ref{fig:CNTpulse}c), corresponding to the fundamental repetition period of the laser cavity, which has an approximate total length of 3.3~m. Mode-beating features in the temporal trace were suppressed by applying a moving average filter to enhance the visualisation of the averaged pulse envelope, shown by yellow plot in Fig.~\ref{fig:CNTpulse}. The standard deviation of the pulse intensity decreased to 3\%. At the same time, the standard deviation of the pulse duration amounted to nearly 1.77\%.

One significant limitation observed during the experiments was thermal damage to the CNT saturable absorber at pump powers exceeding 1.85~mW. This leads to performance deterioration, including substantial instability of the generated pulses. Such degradation in the generation regime was also attributed to the damage of the fluorozirconate fibre tip, which experienced melting and deformation under laser irradiation at these pump power levels. The local heating of the bare fibre facet is caused by OH diffusion from the atmosphere~\cite{caron2012understanding} and heat transfer from the laser-generated beam through the CNT saturable absorber. It is well-known that CNTs demonstrate high heat conductivity~\cite{chernysheva2018revealing}, which increases with the tube concentration or their bundling. As a result, they have found applications as “heat sink” in electronics~\cite{fujii2005measuring}. The application of protective coatings, such as silicon nitride (Si$_3$N$_4$) or aluminum oxide (Al$_2$O$_3$) as fibre endcaps, can help mitigate local heating effects and significantly enhance the long-term stability of fibres in high-power mid-infrared laser systems~\cite{aydin2019endcapping}.

\subsection{Passive Q-switching using a GaSb-based SESAM}

In the next experiment, the laser setup maintained the all-fibre architecture and components from the previous configuration, with a GaSb-based SESAM replacing the gold mirror was replaced. This SESAM was designed for operation at 2.8~$\mu$m and featured a linear reflectivity of approximately 85\%. 

\begin{figure}[!]
\centering
  \includegraphics[width=0.98\linewidth]{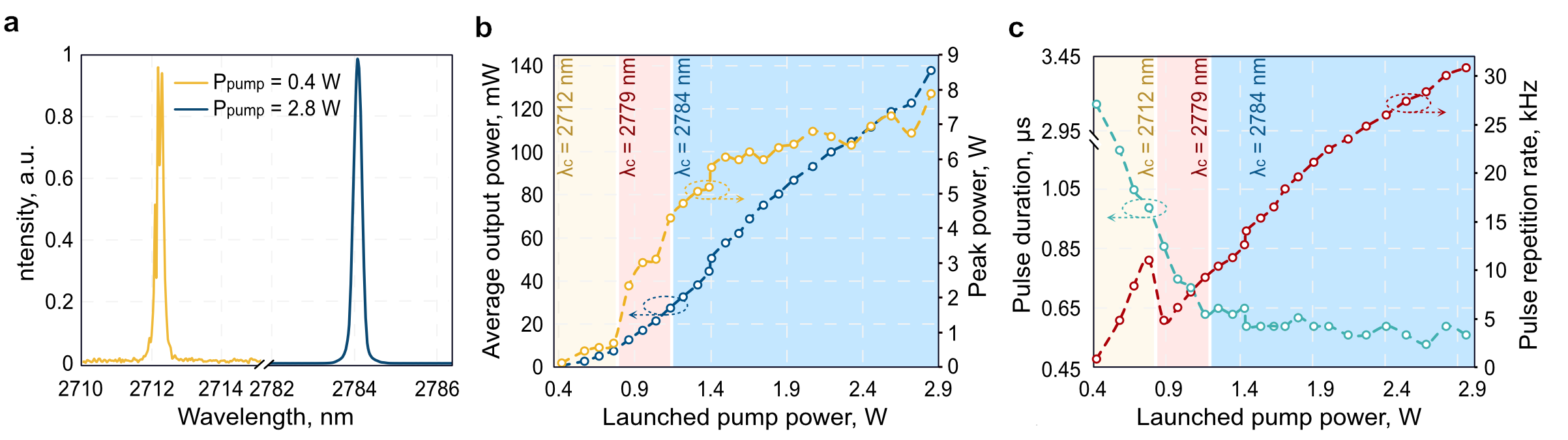}
  \caption{Output parameters of SESAM Q-switched Er-doped ZBLAN fibre laser: (a) Normalised output spectra at lower and upper thresholds of Q-switched generation; variations of (b) output average and peak powers and (c) pulse repetition rate and duration with output power.}
  \label{fig:SESAMoutput}
\end{figure}

Stable Q-switched pulse generation at 1.53~kHz repetition rate emerged at the pump power threshold of 0.4~W, exhibiting a central emission wavelength of 2712~nm with a full-width at half-maximum of 0.2~nm, as shown in Fig.~\ref{fig:SESAMoutput}(a). As pump power increased, the pulse duration progressively decreased from 3.14 to a minimum value of 0.56~$\mu$s, while the repetition rate increased to 31.25~kHz at the maximum pump power. The maximum average output power reached 138~mW, corresponding to the pulse energy of 4.42~$\mu$J and peak power of 7.9~W. In contrast to the laser system employing a CNT saturable absorber, the laser operation central wavelength drifts to the longer range, namely to 2784 nm, with the pump power increase. Due to the saturation of the gain at higher pump rates, the effective gain decreases, leading to the laser generation wavelength shift. Such interplay between the gain spectrum of the doped fibre and the saturation effects manifests through variation of slope laser efficiency and pulse repetition rate change, as shown in Fig.~\ref{fig:SESAMoutput}(b,c). Such gain dynamics were not observed in the previous case of the laser Q-switched using CNT saturable absorber due to the significantly lower range of operated pump power. The overall laser slope efficiency was approximately 6\%. 

\begin{figure}[!]
\centering
  \includegraphics[width=0.48\linewidth]{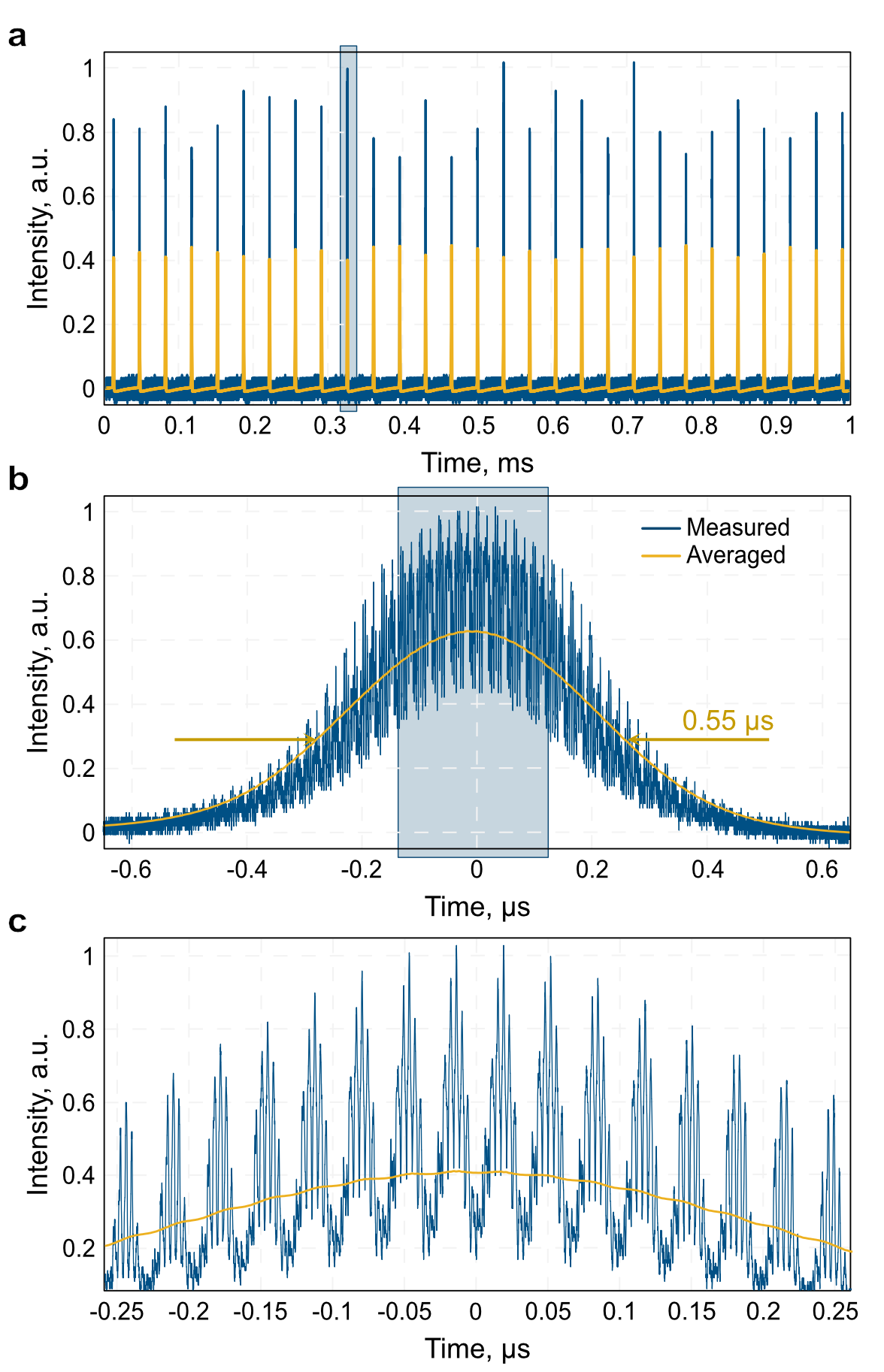}
  \caption{Characteristic output pulse train of SESAM Q-switched Er-doped ZBLAN fibre laser: measured (blue plot) and averaged (yellow plot) (a) at 1-ms time span; (b) zoomed-in individual Q-switched pulse; (c) Zoomed-in oscillations at laser fundamental mode-locking repetition rate. }
  \label{fig:SESAMpulse}
\end{figure}

Similarly to the case of laser Q-switched using a CNT saturable absorber, the oscilloscope traces revealed intensity modulations within the pulse train, shown in Fig.~\ref{fig:SESAMpulse}(a) with a standard deviation of 8.97\%. With the applied moving average filter, the pulse profiles smoothed, reaching a standard intensity deviation of nearly 3\% and 1.06\% standard deviation of pulse duration, indicating slightly improved pulse stability compared to the CNT-based Q-switched laser. Figure~\ref{fig:SESAMpulse}(c) shows mode-locking-like multi-pulsed oscillations with the same periodicity of 32.7~ns as in the case of CNT-Q-switched laser system, corresponding to the fundamental round-trip time of the laser cavity. 

\section{Conclusion}

We have demonstrated Q-switched mode-locking in two identical configurations of all-fibre Er-doped ZBLAN lasers employing either a SESAM or CNT-based saturable absorber. Although both saturable absorbers have been tailored to the 2.8~$\mu$m operation band, the pure mode-locking could not be achieved within the available range of pump power before the degradation of the saturable absorber or unprotected ZBLAN fibre tip was observed. We attribute complex generation dynamics to the insufficient modulation depth of saturable absorbers at the laser operation wavelength. Employing the CNT saturable absorber, the laser generated pulses with a minimum duration of 1.32~$\mu$m and maximum energy up to 1.4~$\mu$J, corresponding to 1.06-W peak power, for 63.1~mW average output power. Alternatively, the application of SESAM in the laser cavity resulted in 560-ns pulse duration at 4.4~$\mu$J energies, corresponding to 7.9-W peak power for 138-mW average power.

The fair comparison of the saturable absorbers has to consider other factors, such as fabrication, long-term stability and robustness, as well as the possibility to tailor further parameters to improve the generation regime. Regarding the fabrication, CNT-based saturable absorbers offer a more straightforward manufacturing process, while the SESAM fabrication is costly and requires cleanroom facilities. Yet, the thermal damage threshold of CNT saturable absorber imposes a limitation on laser output power. For instance, the irreversible damage was observed at an optical fluence of 0.51~J/cm$^2$ in the Q-switched regime, whereas SESAMs withstood fluences of 1.6~J/cm$^2$ or higher. The degradation in the case of the SESAM-based laser was observed at the ZBLAN fibre tip adjacent to the SESAM due to atmospheric vapour absorption rather than the saturable absorber itself. 

The CNT Q-switched pulse demonstrates much more pronounced and deeper modulation with mode-locking-like pulses. At the same time, the mode-locked pulses in the SESAM-based laser system appear more uniform within the Q-switched pulse, indicating a higher pulse train consistency. Such pulse dynamics also manifests a better balance of laser gain depletion and recovery, which is beneficial for stabilising mode-locking generation. The most significant advantage of the CNT saturable absorber, linked to the ease of its fabrication, is the straightforward possibility of optimisation of the saturable absorption parameters by depositing samples with different thicknesses. Thus, the modulation depth and saturation intensity can be tailored to reach the required generation regime.

Overall, the demonstrated Q-switched mode-locking dynamics shows the high potential of both Ga-Sb SESAM and aerosol chemical vapour deposition synthesised CNT to shape the generation of ultrashort pulsed. However, both saturable absorbers emphasize the need for further optimisation to enable stable pure mode-locking at mid-IR wavelength range. These developments will be instrumental in pushing the limits of mid-IR fibre laser technology, paving the way for next-generation compact and efficient ultrafast sources for a broad range of medical, industrial and scientific applications. 

 \begin{figure}[!t]
 \centering
  \includegraphics[width=0.48\linewidth]{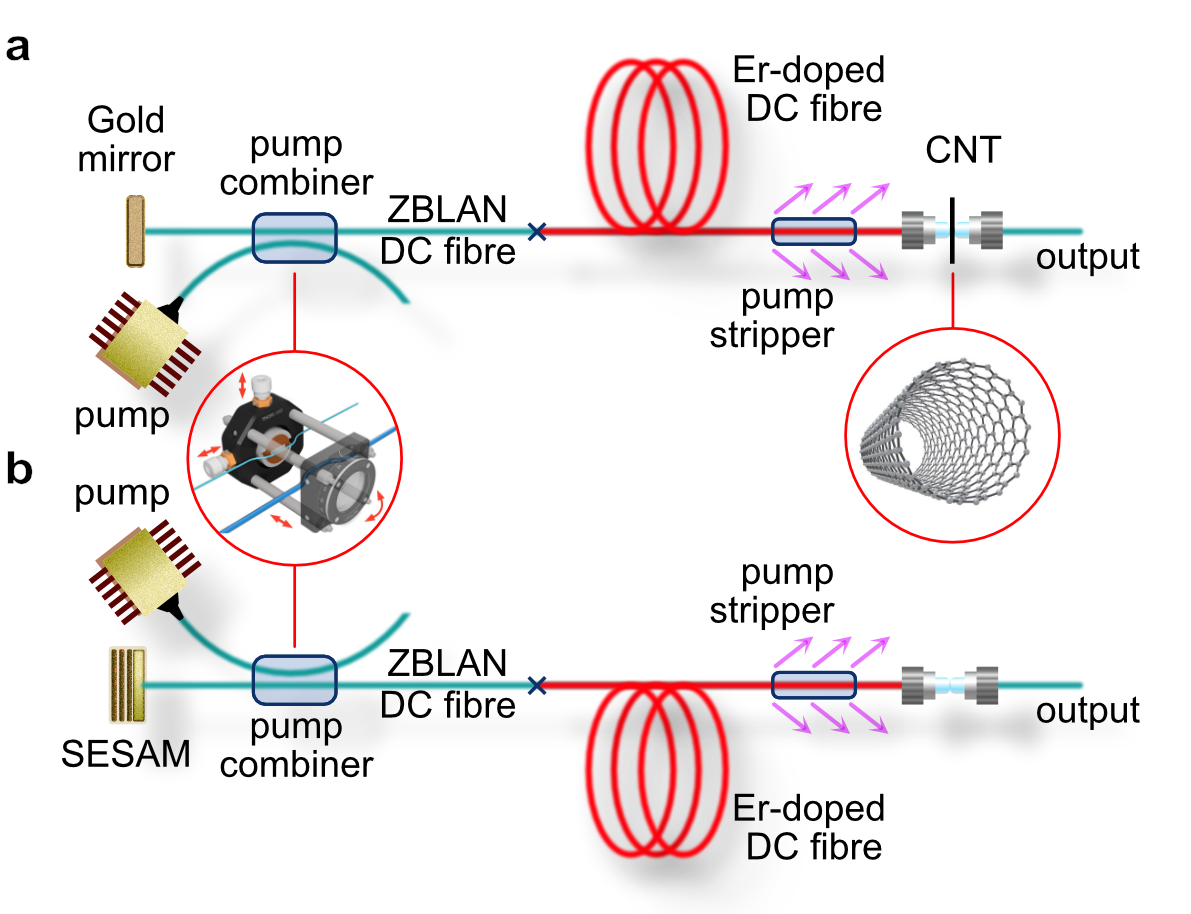}
  \caption{Schematic setups of Er-doped ZBLAN fibre laser systems Q-switched using a) CNTs and b) SESAM}
  \label{fig:setup}
\end{figure}

\section{Experimental Section}
\subsection{Laser system}

The schematics of the Er$^{3+}$-doped ZBLAN fibre laser setups are shown in Fig.~\ref{fig:setup}. In both cases, the pump light at 976~nm is launched into the laser cavity through a side-polished multimode silica-fluoride fibre pump combiner, enabling stable coupling efficiency exceeding 80\% at 980~nm~\cite{perminov2024side}. The active Erbium-doped (7 mol.\%) double-clad ZBLAN fibre (from Le Verre Fluor\'e) has a length of 2.9~m. The core diameter of 15 $\mu$m, a numerical aperture (NA) of 0.12 ensured single-mode light propagation at the generation wavelength. A double-D shaped first cladding with 240 $\times$ 260 $\mu$m diameters and NA of 0.46 allows enhancing pump absorption efficiency when combined with a side-pumping scheme. The residual pump light is extracted through a cladding mode stripper, produced by stripping the outer polymer cladding of the active fibre rear end and applying gel with a refractive index of 1.52. The laser cavity is formed using the Fresnel reflection (4\%) from the FC/PC polished fibre tip, which is also used as the laser output.

The laser system exploits two saturable absorbers. In particular, CNT film was deposited using the dry transfer method onto the output FC/PC polished fibre ferrule. The cavity was formed by placing the gold mirror on the other side. Alternatively, SESAM (from RefleKron Ltd.) was used as a laser cavity without a CNT-based saturable absorber, as shown in Fig.~\ref{fig:setup}.
 
\subsection{Carbon nanotube saturable absorber fabrication and characterisation}

\begin{figure}[!]
\centering
  \includegraphics[width=0.98\linewidth]{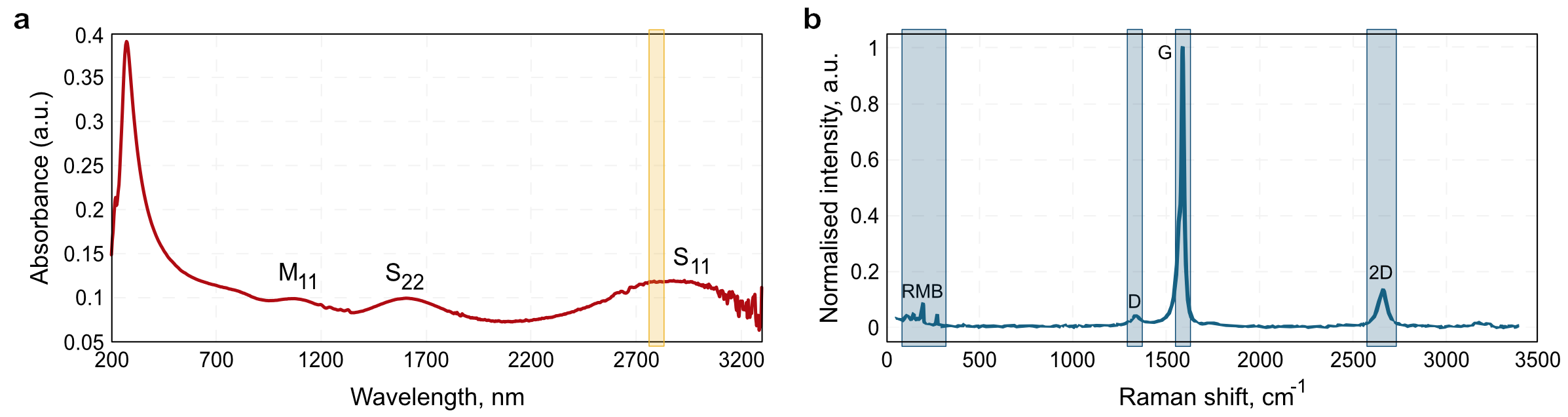}
  \caption{Parameters of CNT saturable absorber: a) Optical absorption spectrum with S$_11$ aligned to laser emission 2.8~$\mu$m wavelength; (b) Raman scattering spectrum showing defectiveness (I$_G$/I$_D$ ratio~=~30) of the CNTs.}
  \label{fig:CNT}
\end{figure}

Single-walled carbon nanotubes were synthesised using the aerosol chemical vapour deposition method, employing ferrocene as the catalyst precursor and a hybrid carbon source comprising ethylene and toluene~\cite{khabushev2022joint}. Ethylene plays a crucial role in activating the Fe catalyst, influencing the diameter of the synthesised CNTs and consequently shifting the S$_{11}$ transition energy to align with the laser emission wavelength of 2.8~$\mu$m, as illustrated in Fig.~\ref{fig:CNT}(a). The mean nanotube diameter was 1.8~nm. While small amounts of ethylene enhance catalyst activation and accelerate nanotube growth, excessive concentrations can lead to catalyst poisoning, adversely affecting synthesis productivity and film quality. In contrast, toluene presents a more stable molecule, which moderates catalyst activation, resulting in a lower CNT yield. When used as the sole carbon source, toluene exhibits limited ability to activate the Fe-based catalyst. However, its slower decomposition kinetics promote enhanced nanotube crystallinity by reducing defect density. This effect is evident in the Raman scattering spectrum (Fig.~\ref{fig:CNT}b). In this spectrum, the I$_G$ mode represents the in-plane stretching of C–C bonds within the nanotube lattice. In contrast, the I$_D$ mode, typically prohibited by Raman spectroscopy selection rules, emerges only when structural defects disrupt the ideal sp$^2$ carbon framework. As depicted in Fig.~\ref{fig:CNT}(b), an I$_G$/I$_D$ ratio reaches 30 when the nanotube diameter corresponds to an S$_{11}$ transition in resonance with the 2.8~$\mu$m wavelength. Such a high ratio confirms the high quality of carbon nanotubes, even with such large diameters. 

After precise tuning the synthesis parameters, CNTs were deposited onto a nitrocellulose filter, forming a thin film of randomly oriented nanotubes with a thickness of 30 nm. The film thickness was controlled by adjusting the collection time. The dry transfer technique was employed for sample preparation to encapsulate the CNT saturable absorber between two FC/PC ferrules~\cite{kaskela2010aerosol,nasibulin2011multifunctional,mkrtchyan2019dry}.

\medskip

\medskip
\textbf{Acknowledgements} \par %delete if not applicable))
The work has been funded by the German Federal Ministry of Education and Research (BMBF) and supervised by the VDI Technology Center under number 13N15464 "Leibniz Center for Photonics in Infection Research (LPI): Multidimensional, multimodal, intelligent imaging platforms". The authors also acknowledge the support from the European Regional Development Fund. The work was partially supported by the Russian Science Foundation grant No. 20-73-10256 (synthesis of a SWCNT film with a predefined diameter distribution).

\medskip
\textbf{Conflict of Interest}\par
The authors declare no conflict of interest.

\medskip
\textbf{Data Availability Statement}\par
The data that support the findings of this study are openly available in figshare at\\ http://doi.org/10.6084/m9.figshare.28477127, reference number 28477127.
% References

\bibliographystyle{MSP}
\bibliography{sample}

\medskip

% Figures/tables and captions
% Permission statements are required for all figures reproduced or adapted from previously published articles/sources. Please also ensure that all necessary permissions to reproduce images have been received
% Please remove these statements for original figures

% Table of contents entry should be 50 - 60 words long
% Image should be 55 mm broad and 50 mm high or 110 mm broad and 20 mm high

\begin{figure}[bp]
\textbf{Table of Contents}\\
\medskip
  \includegraphics{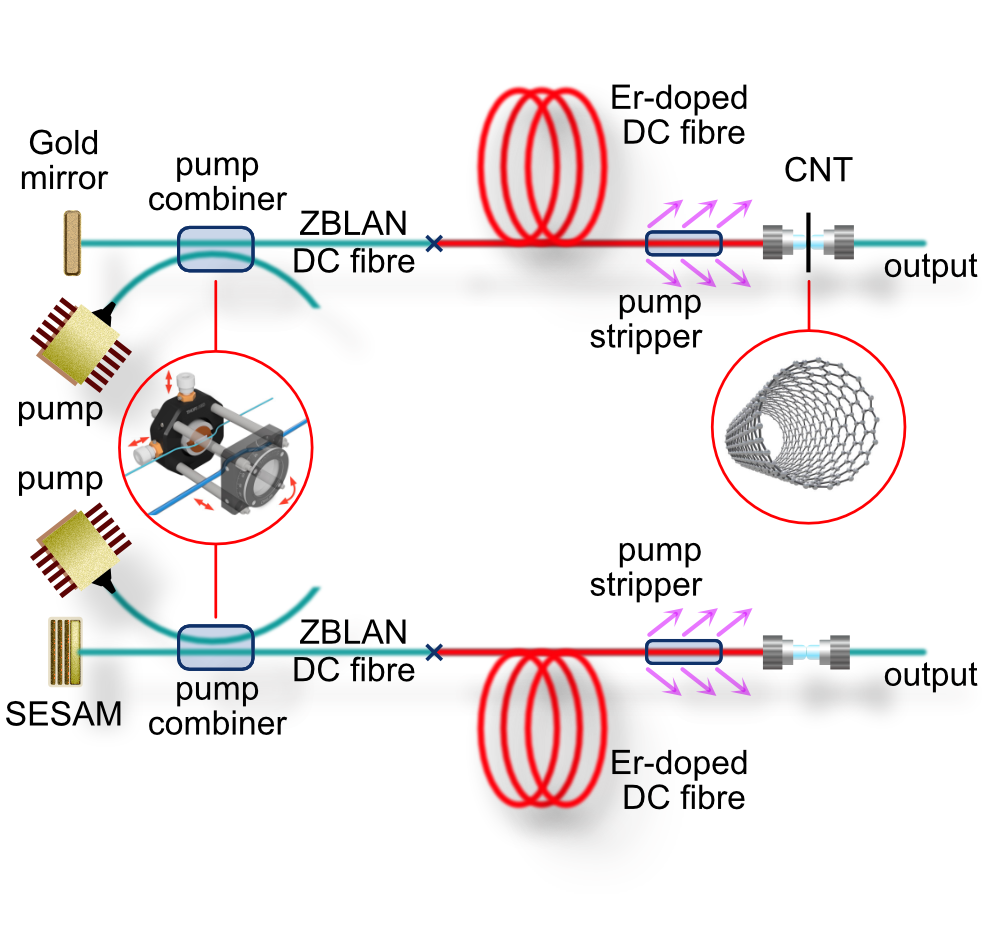}
  \medskip
  \caption*{We compare Q-switched mode-locking in Er-doped ZBLAN fibre lasers using an aerosol-synthesised carbon nanotube saturable absorber with a GaSb-based SESAM. While SESAM enables shorter pulses with higher peak power, CNT absorbers offer tunability and cost-effectiveness. These findings advance mid-IR ultrafast laser development, paving the way for compact, high-power sources for sensing, medical, and industrial applications.}
\end{figure}

\end{document}